\documentclass[a4paper,fleqn,usenatbib]{mnras}

\usepackage{newtxtext,newtxmath}

\usepackage[T1]{fontenc}
\usepackage{ae,aecompl}

\usepackage{graphicx}	
\usepackage{amsmath}	
\usepackage{amssymb}	

\usepackage{color}
\usepackage{ulem}

\title[Vertical amplitudes of disk-crossing orbits]{
Envelopes and vertical amplitudes of disk-crossing orbits
}

\author[R. S. S. Vieira and J. Ramos-Caro]{
Ronaldo S. S. Vieira$^{1}$\thanks{E-mail: rssvieira@ime.unicamp.br (RSSV)}
and Javier Ramos-Caro$^{2}$
\\
$^{1}$Department of Applied Mathematics, University of Campinas, 13083-859 Campinas, SP, Brazil\\
$^{2}$Departamento de F\'{i}sica, Universidade Federal de S\~{a}o Carlos, 13565-905, S\~{a}o Carlos, SP, Brazil
}

\date{Accepted XXX. Received YYY; in original form ZZZ}

\pubyear{2019}

\begin{document}
\label{firstpage}
\pagerange{\pageref{firstpage}--\pageref{lastpage}}
\maketitle

\begin{abstract}
We recently found that regular orbits in axially symmetric galactic disks have their envelopes $Z(R)$ accurately described by the relation
$Z(R)\propto[\Sigma_I(R)]^{-1/3}$, if their amplitudes are comparable to the disk thickness, 
where $\Sigma_I$ is the surface density of the disk (integrated over its whole vertical range). Moreover, the usual adiabatic approximation gives a good description of the orbits' envelopes for low vertical amplitudes. However, these two approaches are apparently disconnected, since their expressions differ qualitatively. 
Our purpose in this paper is to fill this gap  by extending these previous formulae to regular orbits with arbitrary vertical amplitudes inside the disk.
We compare existing $Z(R)$ estimates: the razor-thin disk case, the adiabatic approximation (low-amplitude orbits in three-dimensional disks), and the integrated surface-density estimate (high-amplitude orbits in three-dimensional disks) in order to establish a connection between them.
The formula presented here links the aforementioned results in an elegant and continuous way, being valid for vertical amplitudes throughout the whole vertical extension of the disk and with an expression which has the same form for all regimes.
The advantage of the present formalism is the dependence of $Z(R)$ only on  observable quantities, namely the disk's vertically integrated surface density, without the need to obtain the gravitational potential for the system.
\end{abstract}




\section{Introduction}

It is widely accepted nowadays that general Hamiltonian dynamical systems have their phase space composed of both regular and chaotic regions. This fact was confirmed by a variety of numerical experiments
\citep[e.g. ][]{henonheiles1964AJ,contopoulosOCDA2002,hunter2003LNP,
hunter2005NYASA,
ramoscaroLopezsuspesGonzalez2008MNRAS}.
Seminal studies showed this pattern decades ago in galactic dynamics
\citep[][see \citealt{contopoulosOCDA2002} for a thorough review]{contopoulos1960ZA, ollongren1962BAIN, contopoulos1963AJ, henonheiles1964AJ}, 
starting the ``quest'' for the third integral of motion $I_3$ in axially symmetric disk systems. Integrability by means of
polynomial invariants was studied in \citet{hietarinta1987PhR}, 
whereas St\"ackel potentials \citep{kuzmin1952PTarO, kuzmin1956AZh, dezeeuw1985MNRAS,dezeeuw1988NYASA,
dejongheDezeeuw1988ApJ,dezeeuwHunter1990ApJ} became the standard example
of integrable galactic models. 

The current theoretical understanding of the third-integral formalism in axially symmetric disk galaxies is based on the existence of adiabatic invariants 
around equatorial circular orbits \citep{binneytremaineGD, binney2010MNRAS, binneyMcmillan2011MNRAS}, as well as on the order-by-order expansion
of the invariant \citep{contopoulos1960ZA,contopoulos1963AJ,contopoulosOCDA2002}.
There was also a recent attention devoted to the numerical obtention of action-angle coordinates for nearly integrable regions in phase space
\citep{mcmillanBinney2008MNRAS, binney2012MNRASactions, sanders2012MNRAS, sandersBinney2014MNRAS},
as well as to efficient methods of obtaining polynomial approximate integrals \citep{bienaymeTraven2013AA} and closed-formula expressions for $I_3$ \citep{bienaymeRobinFamaey2015AA,vieiraRamosCaro2016CeMDA}. These many complementary approaches
improved the development of dynamical models for the Galaxy 
\citep{dezeeuw1987IAUS,binney2010MNRAS,binneyMcmillan2011MNRAS,
binney2012MNRASdynamical}, which 
are a key ingredient to analyze and interpret the new kinematic data obtained from Galaxy surveys \citep{binneySanders2014IAUS}, such as \textit{Gaia} DR2 \citep{Gaia2018AA}.

Let us assume an axially symmetric system. An approximate third integral gives us, in particular, the form of the orbit's amplitude $Z$ as a function of radius $R$ on the $R$--$z$ plane (its ``envelope" $Z(R)$). In spite of the significant advances in recent years, closed-form analytical approximations for the non-classical integral $I_3$ (and for the corresponding envelope)
are still very scarce, and the existent results have limited range of validity. The usual adiabatic approximation based on 
decoupled radial and vertical motions
\citep{binneytremaineGD}, for instance, is only valid very close to the equatorial plane \citep{binneyMcmillan2011MNRAS,vieiraRamoscaro2014ApJ}. 
It was recently found,
based on the study of orbits in razor-thin disks \citep{vieiraRamoscaro2015MG13,vieiraRamosCaro2016CeMDA},
that a simple relation can describe the envelopes $Z(R)$ of disk-crossing orbits whose amplitudes
are near the disk's vertical edge \citep{vieiraRamoscaro2014ApJ}. The orbit's amplitude scales with a fixed power of 
the integrated surface density. 
However, the formula fails for intermediate and low vertical excursions.
In summary, the two above limits, with near-equatorial and near-edge vertical amplitudes, can be understood in terms of 
straightforward physical arguments. But, in order to have a quantitative physical description of the behaviour of off-equatorial orbits with
varying vertical amplitudes, a connection between these two regimes must be done. 

The purpose of this paper is to fill this gap. We present an extension of the previous formulae, which connects the two regimes described above
in an unified and elegant way. 
We obtain an expression which describes the envelopes $Z(R)$ of disk-crossing (tube) orbits with 
vertical amplitudes spanning the whole vertical extension of the disk, 
and which gives the aforementioned known approximations in their respective ranges of validity.
Moreover, this expression for $Z(R)$ depends only on the disk's dynamical surface density, integrated up to the orbit's vertical amplitude (see Eqs.~(\ref{eq:Sigma-Z})--(\ref{eq:Z-Sigma})). In this way, the density profile completely determines the envelopes of tube orbits for any vertical range of amplitudes inside the disk.

The paper is structured as follows: Section \ref{sec:envelopes} summarizes the two known cases which give good predictions for the envelopes $Z(R)$: the adiabatic approximation (\ref{eq:Z-AA}) (orbits with very small vertical amplitudes) and the surface-density estimate (\ref{eq:Z-zeta}) for the envelopes of orbits near the disk edge. It also presents the unified framework which connects the adiabatic approximation and the near-edge estimate via an elegant procedure, based solely on the disk's surface density (integrated up to the orbit's amplitude). Section \ref{sec:iterative} briefly comments on the method of successive approximations used to numerically calculate the envelopes given by the integral equation resulting from our new framework, and  Section \ref{sec:numerical} presents numerical results which confirm our predictions. 
We present our conclusions in Section \ref{sec:conclusions}.

\section{$Z(R)$ envelopes on the $R$--$z$ plane}
\label{sec:envelopes}

We work in cylindrical coordinates ($R,z,\varphi$) and consider axially symmetric disks with a density profile $\rho(R,z)$ concentrated around the equatorial plane $z=0$.
As mentioned in the Introduction, there are different approaches to estimate the orbits' envelopes on the meridional plane $R$--$z$. 
We know that the epicyclic approximation gives us an adiabatic invariant of the form (when evaluated at the envelope $Z(R)$)
  \begin{equation}\label{eq:I3-AA}
   I_3^{\ (AA)}=Z(R)\big[\Phi_{zz}(R,0)\big]^{1/4}
  \end{equation}
as approximate third integral for the system \citep{binneytremaineGD}, where $\Phi$ is the gravitational potential, $\Phi_{zz}=\partial^2\Phi/\partial z^2$,  and $Z$ is the orbit's vertical amplitude.
This expression results in \citep{binneytremaineGD}
  \begin{equation}\label{eq:Z-AA}
   Z(R)\propto\big[\Phi_{zz}(R,0)\big]^{-1/4}
  \end{equation}
for the vertical amplitudes of nearly equatorial orbits.
However, it is well known that this approximation 
has a very limited range of applicability, being valid only for orbits whose vertical excursions are very small
\citep{binneyMcmillan2011MNRAS}. In particular, their vertical amplitudes must be much smaller than the disk thickness
\citep{vieiraRamoscaro2014ApJ}. Variants of this formula were proposed in order to overcome this problem, 
as for instance the corrections to the adiabatic approximation of \citet{binneyMcmillan2011MNRAS}. In spite of the fact that this correction has a wider range of validity and 
can describe accurately the corresponding off-equatorial orbits, it introduces additional phenomenological parameters 
which must be adjusted \textit{a posteriori} \citep[][see also \citealt{vieiraRamoscaro2014ApJ}]{binneyMcmillan2011MNRAS}.

On the other hand, \citet{vieiraRamoscaro2014ApJ} obtained a simple expression for the envelopes of disk-crossing
orbits whose vertical amplitudes are comparable to the disk thickness $\zeta$:
  \begin{equation}\label{eq:Z-zeta}
   Z(R)\propto\big[\Sigma_I(R)\big]^{-1/3}
  \end{equation}
where 
  \begin{equation}\label{eq:Sigma-I}
  \Sigma_I(R)=\int_{-\zeta}^{\zeta} \rho(R,z)\, dz,
  \end{equation}
and $\rho$ is the system's density distribution (which may contain different disk components or even a sparse dark-matter halo).
This formula was based on our results for razor-thin disks \citep{vieiraRamoscaro2015MG13,vieiraRamosCaro2016CeMDA} and depends only on the integrated surface density $\Sigma_I$, which we call for brevity ``disk's surface density''. Therefore, Eq.~(\ref{eq:Z-zeta}) can be evaluated from observational data alone (without the knowledge of the gravitational potential). 
However, although we know that expressions (\ref{eq:Z-AA}) and (\ref{eq:Z-zeta}) are both valid in their respective
ranges of applicability, there is a variety of disk-crossing orbits which ``live'' in the intermediate regions, that is, whose
vertical amplitudes are spread throughout the range $0<Z<\zeta$. These orbits cannot be described by neither of the expressions 
(\ref{eq:Z-AA}) or (\ref{eq:Z-zeta}), since the assumptions which led to each of these equations cease to be valid for intermediate-amplitude orbits.

\subsection{Connecting the formulae}\label{sec:connectinng}

In order to keep the framework of writing the orbits' envelopes for arbitrary vertical amplitudes only in terms of the disk density, we must find a link between expressions (\ref{eq:Z-AA}) and (\ref{eq:Z-zeta}).
Doing this in an unified way will allow us to obtain an expression 
which is valid for the whole range of vertical amplitudes inside the disk, and which has the correct limits for almost-zero
and near-edge amplitudes. The connection is done once we note that Poisson equation can be written in the form
\citep{binneytremaineGD}
  \begin{equation}
   \nabla^2\Phi=\frac{\partial^2\Phi}{\partial z^2}+ \frac{1}{R}\frac{d v_c^2}{dR} = 4\pi G\rho
   \label{eq:poisson}
  \end{equation}
on the equatorial plane, where $v_c$ is the speed of a circular orbit of radius $R$.
Since most disk galaxies have very small velocity gradients far from the central region \citep{sofueRubin2001ARAA}, the term involving $v_c$ may be neglected if 
the density distribution is large enough on the equatorial plane. 
We would like to have
  \begin{equation}\label{eq:Z-maybe}
   Z(R)\propto\big[\Phi_{zz}(R,0)\big]^{-1/4}\propto\big[\Sigma_0(R)\big]^{-1/3}
  \end{equation}
in the domain of validity of the adiabatic approximation (\ref{eq:Z-AA}) (i.e. for very small vertical amplitudes),  
where $\Sigma_0$ is an appropriately defined integrated density. The most natural choice would be 
  \begin{equation}
   \Sigma_0(R)=\int_{-Z(R)}^{Z(R)} \rho(R,z)\, dz.
  \end{equation}
Since $Z(R)$ is small, we may consider the integrand in the above expression as a constant function of $z$ (evaluated at $z=0$),
$\rho(R,z)\approx\rho_0(R)$, which gives us
  \begin{equation}\label{eq:Sigma-0-approx}
   \Sigma_0(R)\approx 2\rho_0(R)Z(R).
  \end{equation}
With the aid of Eq.~(\ref{eq:poisson}), and neglecting the term involving the velocity gradient, we obtain
  \begin{equation}
   \Sigma_0(R)\propto Z(R)\Phi_{zz}(R,0),
  \end{equation}
and thus
  \begin{equation}
   \Phi_{zz}(R,0)\propto\frac{\Sigma_0(R)}{Z(R)}.
  \end{equation}
Substituting in Eq.~(\ref{eq:Z-AA}), we obtain 
  \begin{equation}
   Z^4(R)\propto\frac{1}{\Phi_{zz}(R,0)}\propto\frac{Z(R)}{\Sigma_0(R)}, 
  \end{equation}
that is, 
  \begin{equation}
   Z(R)\propto\big[\Sigma_0(R)\big]^{-1/3}.
  \end{equation}
  
Therefore, Eqs.~(\ref{eq:Z-AA}) and (\ref{eq:Z-zeta}) will have the same form if the ``surface density''
appearing in the formula (\ref{eq:Z-zeta}) for $Z(R)$ is, instead of (\ref{eq:Sigma-I}),
  \begin{equation}\label{eq:Sigma-Z}
   \Sigma(R)=\int_{-Z(R)}^{Z(R)} \rho(R,z)\, dz,
  \end{equation}
where the limit of integration $Z(R)$ is the orbit's vertical amplitude (which depends on the galactocentric radius $R$). Note that, when the orbit's amplitude is comparable to the disk thickness, the two formulae (\ref{eq:Z-zeta}) and (\ref{eq:Z-Sigma}) give practically the same result.
We call the $\Sigma(R)$ given by Eq.~(\ref{eq:Z-Sigma}) the ``orbit's surface density''. 
The orbit's amplitude as a function of $R$ will be given by the implicit equation
  \begin{equation}\label{eq:Z-Sigma}
   \frac{Z(R)}{Z(R')}=\left[\frac{\Sigma(R')}{\Sigma(R)}\right]^{1/3}
  \end{equation}
for $R$, $R'$ in the orbit's radial range,
with $\Sigma$ given as a function of $Z$ by expression (\ref{eq:Sigma-Z}). We claim that this expression 
is valid through the whole vertical range of the disk, that is, for regular orbits with amplitudes in the whole range $0<Z<\zeta$.

We might approximate the vertical amplitude $Z(R)$ appearing in (\ref{eq:Sigma-Z}) 
by an average value along the orbit, if the amplitudes do not vary much.
In this case, Eq.~(\ref{eq:Z-Sigma}) is given explicitly as a function of $R$, once we have an estimate for the vertical amplitude, and the equations would give an explicit formula for $Z(R)$.
However, this procedure can produce additional errors in the prediction for the envelope $Z(R)$, since we neglect the $R$-dependence of the integration interval: 
the ratio $\Sigma/\Sigma_I$ is then more likely to keep a constant value along the orbit. The prediction for $Z(R)$ will
therefore be very similar to Eq.~(\ref{eq:Z-zeta}),
and the correspondence would be exact if the density profile had a separable form, 
$\rho(R,z)=\Sigma_I(R)\xi(z)$, with $\int\xi dz=1$. For flattened disks, 
the aforementioned plane-parallel case is only approximately valid in general, so an important piece of information is being underestimated
if we neglect the $R$-dependence of the integration limits.
Moreover, Eq.~(\ref{eq:Z-zeta}) is only valid for amplitudes which are close to 
the disk's vertical edge \citep{vieiraRamoscaro2014ApJ}, so in the discussion which follows we will always refer to the implicit expression
(\ref{eq:Z-Sigma}), where the integration limits in (\ref{eq:Sigma-Z}) are the actual values of the orbit's envelope.
We test below this expression for a variety of orbits in disk-like potentials, 
with amplitudes varying in the range $0<Z\lesssim\zeta$.

\section{An iterative procedure for $Z(R)$}\label{sec:iterative}

Equation (\ref{eq:Z-Sigma}) does not give us a direct method to obtain $Z(R)$, since the integration limits of $\Sigma$, Eq.~(\ref{eq:Sigma-Z}), are precisely the 
function sought. We would have then to solve an integral equation to obtain the orbit's envelope. 
One way to circumvent this issue is to consider an iterative procedure by means of a Neumann series (the well-known ``method of successive approximations'', see \citealp{arfken1972mathematical}), 
in which a function $Z_o(R)$ is chosen as an initial guess, inserted on the right-hand 
side of the equation for $Z(R)$ (see Eqs.~(\ref{eq:Sigma-Z})--(\ref{eq:Z-Sigma}), with $\pm Z_o(R)$ being the integration limits for $\Sigma$):
	\begin{equation}\label{eq:Z-iterative}
	 Z(R)=A\big[\Sigma(R)\big]^{-1/3},
	\end{equation}
where $A$ is a constant and $\Sigma(R)$ is calculated from Eq.~(\ref{eq:Sigma-Z}). 
This first step will generate a new function $Z_1(R)$, which will be used in
the second step as the new integration limit for $\Sigma$ in Eq.~(\ref{eq:Z-iterative}), giving rise to a function $Z_2(R)$.
After $N$ steps, we will have a sequence of functions 
	\begin{equation}\label{eq:Z-sequence}
	 Z_o\to Z_1\to Z_2\to ...\to Z_N.
	\end{equation}

Ideally, the sequence of functions $(Z_n)$ will be uniformly convergent in the range of radii
relevant for the orbit, and its limit will be the solution $Z(R)$ to 
Eqs.~(\ref{eq:Sigma-Z})--(\ref{eq:Z-Sigma}). In that case, the sequence ($a_n$) given by
	\begin{equation}
	 a_n= \operatorname*{max}_{R\in I}\,\frac{\big|Z_n(R)-Z_{n-1}(R)\big|}{\big|Z_n(R)\big|}
	\end{equation}
will converge to zero, where $I$ is a previously defined interval corresponding to the orbit's radial range. We therefore look for an approximate solution to these equations by obtaining a value of $N$
for which $a_N<\epsilon$, where $\epsilon$ is a tolerance parameter. The iterated function $Z_N$ will therefore be an 
approximation to the actual solution $Z(R)$ within the error $\epsilon$. 
This procedure is, therefore, dependent only on the initial guess $Z_o(R)$ for $Z$; it does not rely upon an \textit{a priori} knowledge of the envelope. In this way, it can be used
to make predictions about the orbit's shape knowing (for instance) only one point on the 
zero-velocity curve. We will follow this procedure to numerically calculate the orbits' envelopes for specific example disk potentials.

\section{Tests with simple disk potentials}
\label{sec:numerical}

We perform in this Section numerical experiments in specific disk-like potentials in order to test the validity of prediction (\ref{eq:Sigma-Z})--(\ref{eq:Z-Sigma}), for a variety of orbits with amplitudes spread in the range $0<Z(R)<\zeta$. We find in all cases that expression (\ref{eq:Z-Sigma}), with $\Sigma(R)$ given by the orbit's surface density (\ref{eq:Sigma-Z}), is a good approximation to the real orbits' envelopes, with errors below 1\%. The prediction (\ref{eq:Z-Sigma}) is better than both the adiabatic approximation (\ref{eq:Z-AA}) and the estimate based on the disk's surface density (\ref{eq:Z-zeta}), even in their ranges of validity (very low vertical amplitudes and amplitudes near the disk edge, respectively). 

We choose as reference the point on the orbit with the largest vertical amplitude $Z$. The predictions for the envelopes given by the different approximations are made to agree with the vertical amplitude of the numerically integrated orbit at this point, and we compare the three formulae (\ref{eq:Z-AA}), (\ref{eq:Z-zeta}), and (\ref{eq:Z-Sigma}) along the orbit's radial range. Equations (\ref{eq:Z-AA}) and (\ref{eq:Z-zeta}) give explicitly $Z$ as a function of $R$ and can be readily compared with the numerically calculated envelope. On the other hand, Eq.~(\ref{eq:Z-Sigma}), with $\Sigma(R)$ given by the orbit's surface density (\ref{eq:Sigma-Z}), does not give $Z(R)$ explicitly. In order to obtain the predicted envelope we apply the iterative method described in Section~\ref{sec:iterative}, starting with $Z_o(R)$ given by the adiabatic approximation (\ref{eq:Z-AA}), $Z_o(R)\propto\big[\Phi_{zz}(R,0)\big]^{-1/4}$. We perform 10 iterations ($N=10$) for each orbit in order to obtain an accurate approximation to Eqs.~(\ref{eq:Sigma-Z})--(\ref{eq:Z-Sigma}). Very good approximations already appear for smaller $N$.





\subsection{Miyamoto-Nagai disk}

We consider as the main example the Miyamoto-Nagai potential \citep{miyamotoNagai1975PASJ,binneytremaineGD}
  \begin{equation}\label{eq:Phi-MN}
   \Phi=-\frac{GM}{\sqrt{R^2+(a+(z^2+b^2)^{1/2})^2}},
  \end{equation}
which is commonly used to model the disk component of Galaxy mass models \citep{allenSantillan1991RMxAA,irrgangEtal2013AA, barrosLepineDias2016AA} 
and also to analyze the Saggitarius dwarf galaxy debris
\citep{johnstonEtal1995,helmi2004MNRAS,lawEtal2009ApJL,
ibataEtal2013ApJL,degWidrow2013MNRAS}. The disk's vertical edge is estimated as $\zeta/a\approx 3b/a$ if $b\ll a$ \citep{vieiraRamoscaro2014ApJ}. We analyze various orbits for different values of energy $E$, angular momentum $L_z$, and initial conditions. 
We consider only the cases with a sufficiently flattened density profile ($b/a \lesssim 1/10$).

\begin{figure*}
\includegraphics[width=\columnwidth]{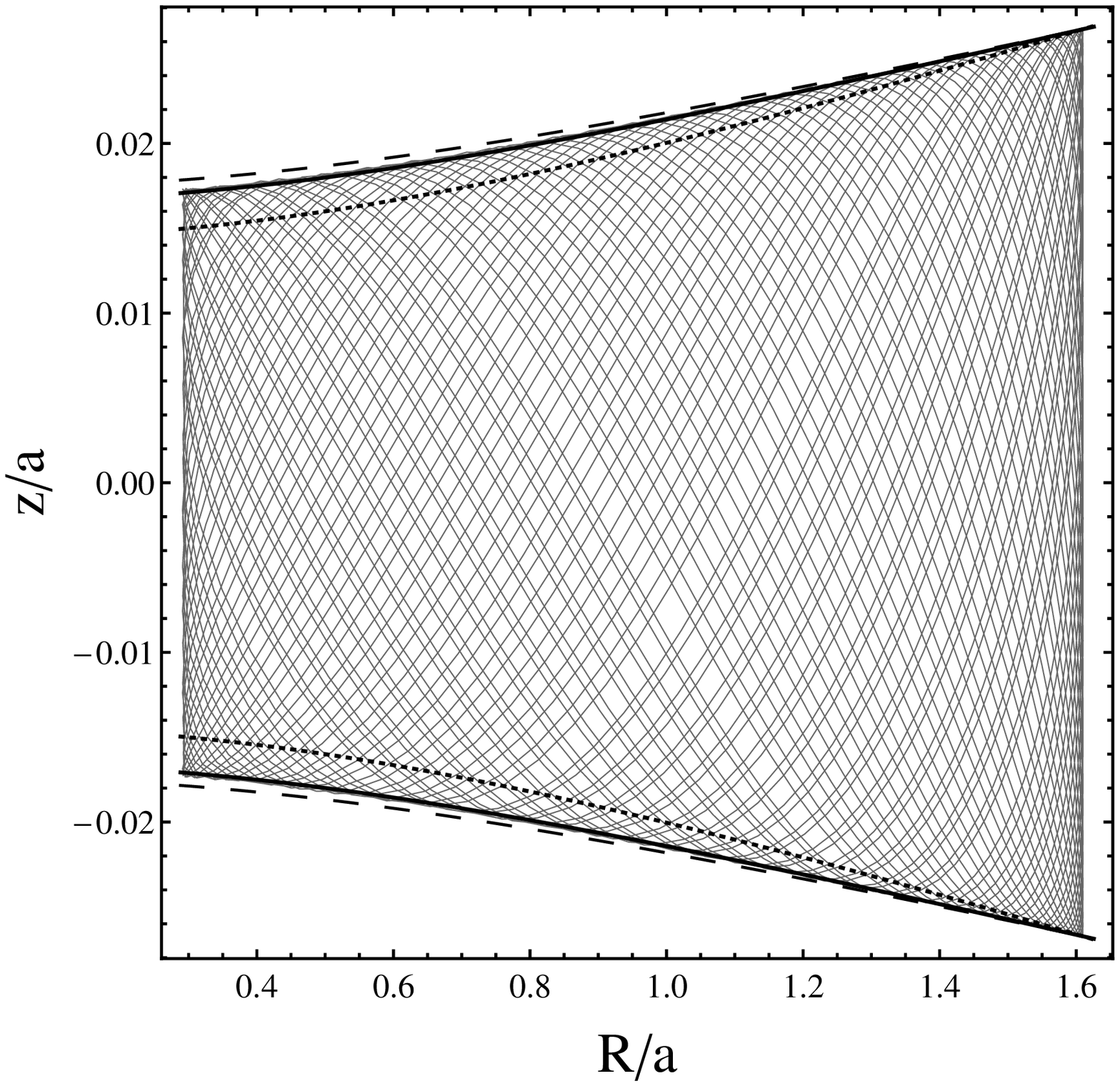}
\includegraphics[width=\columnwidth]{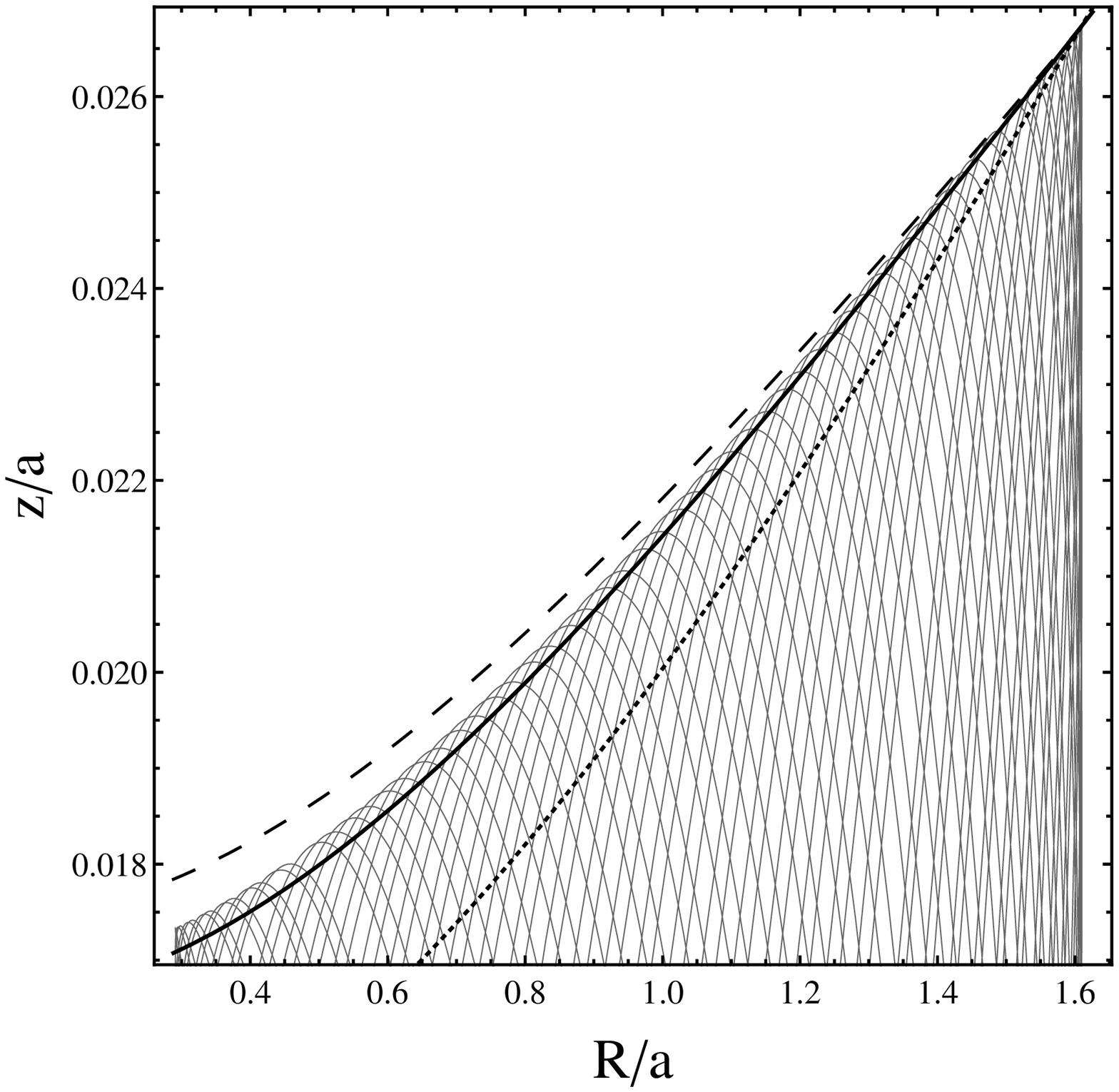}
 \caption{Left: Orbit in the Miyamoto-Nagai disk with $b/a=1/10$ (we have $\zeta/a\approx0.3$). Here, $L_z/\sqrt{GMa}=0.8$ and $Ea/GM=-0.050$.  
 Moreover, $R_0/a=1.61$, $z_0/a=1.1\times10^{-2}$, $P_{R0}=0$.  The orbit is in the range of validity of the adiabatic approximation ($Z\ll\zeta$).
 The dashed curve corresponds to the adiabatic approximation (\ref{eq:Z-AA}), the dotted curve corresponds to the estimate via the disk's surface density (\ref{eq:Z-zeta}), and the solid curve to the estimate via the orbit's surface density (\ref{eq:Sigma-Z})--(\ref{eq:Z-Sigma}) calculated by performing 10 successive approximations starting with the adiabatic approximation (\ref{eq:Z-AA}) (see Section \ref{sec:iterative}). 
 Right: Zoom of the upper part of the orbit, showing its envelope. The curve (\ref{eq:Z-Sigma}) lies between the other two curves, being the best approximation among the three formulae to describe the orbit's envelope. 
 In particular, it is evident from this panel that the solid curve (\ref{eq:Z-Sigma}) improves the usual adiabatic approximation (in its range of validity), since it considers different vertical ranges of integration for each value of $R$.
 }
 \label{fig:teste1}
\end{figure*}

\begin{figure*}
\includegraphics[width=\columnwidth]{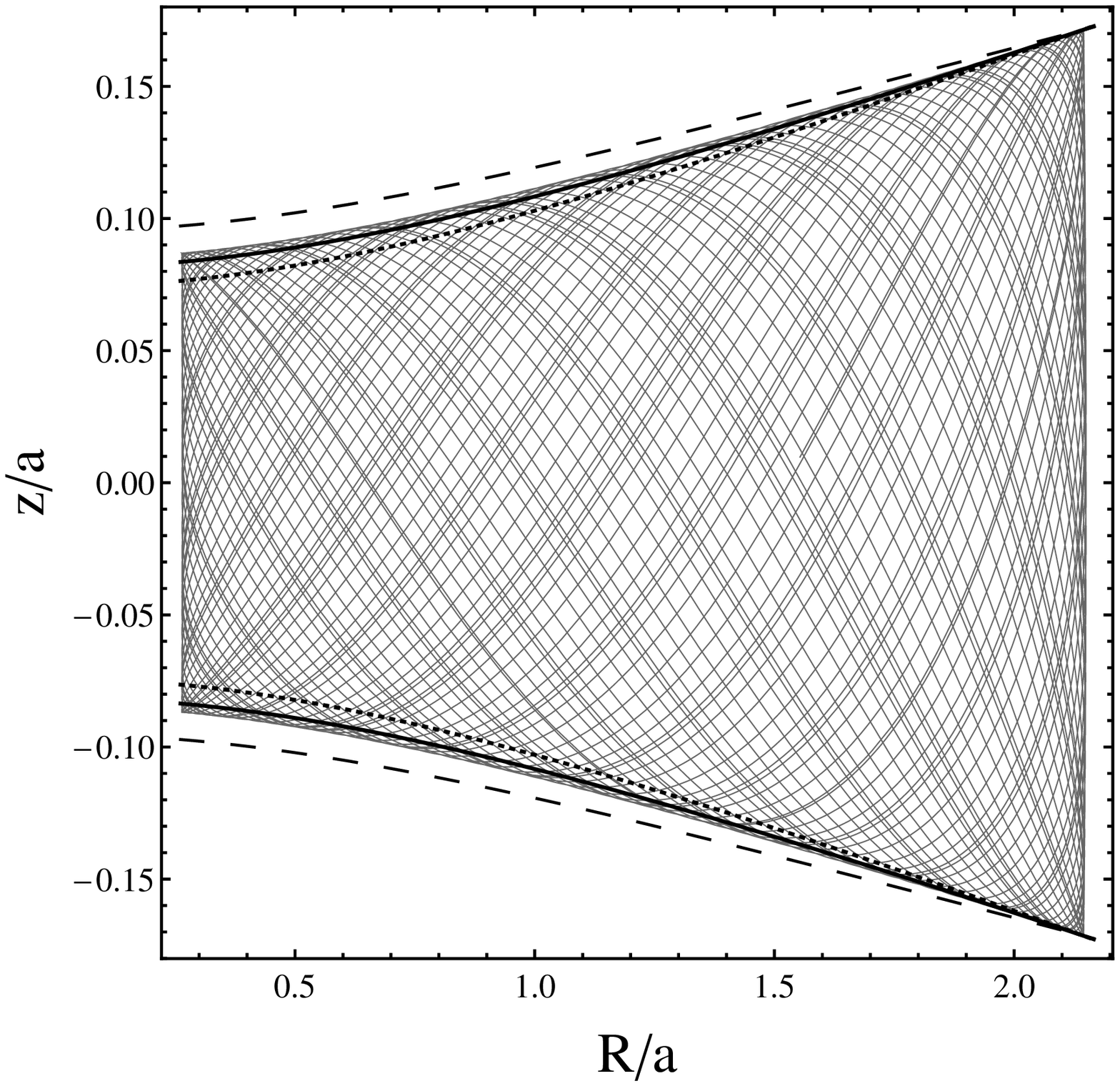}
\includegraphics[width=\columnwidth]{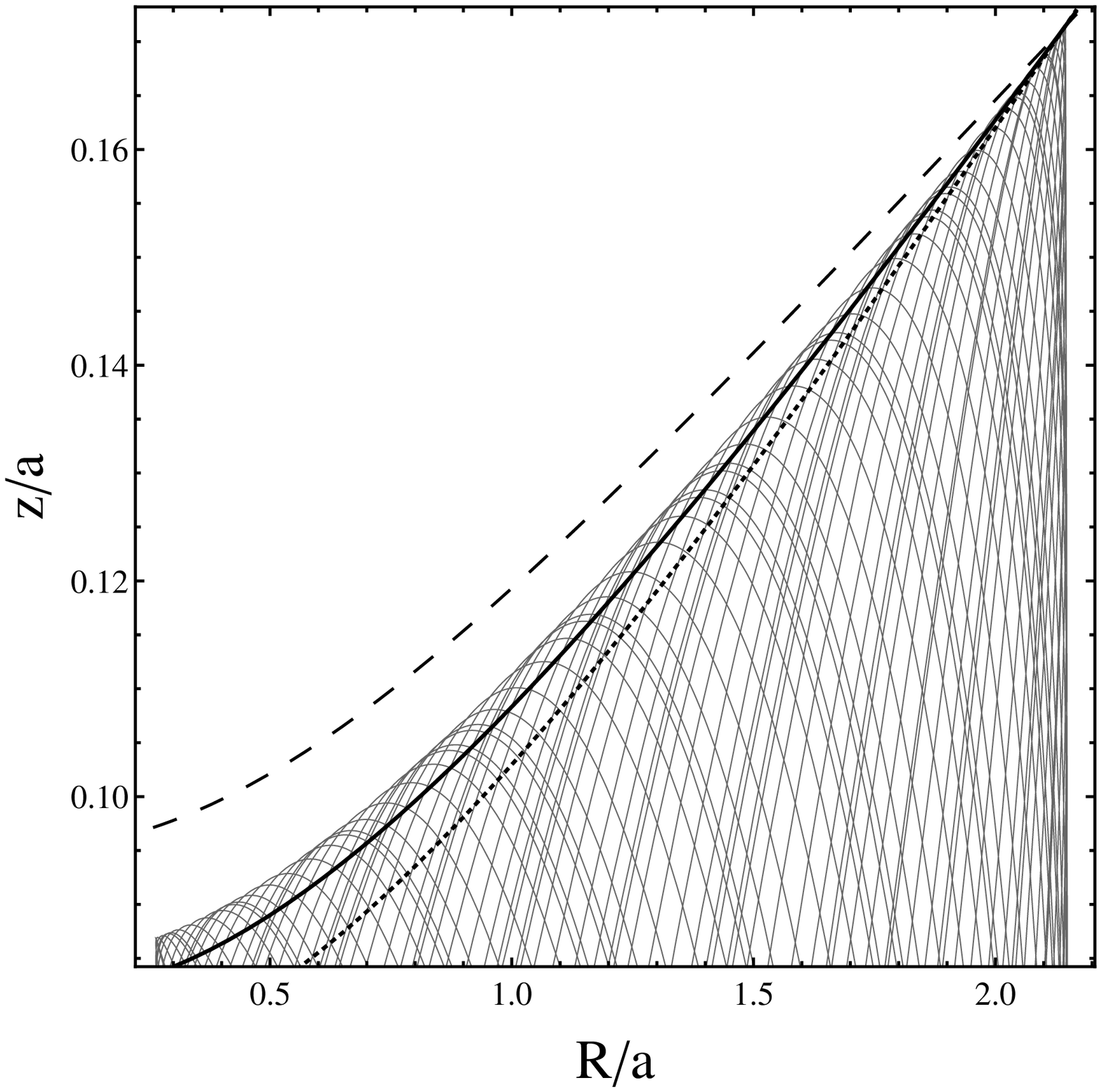}
 \caption{Left: Orbit in the Miyamoto-Nagai disk with $b/a=1/10$ (we have $\zeta/a\approx0.3$). Here, $L_z/\sqrt{GMa}=0.8$ and $Ea/GM=-0.040$. 
 Moreover, $R_0/a=2.15$, $z_0/a=1.1\times10^{-2}$, $P_{R0}=0$. Labels are as in Fig. \ref{fig:teste1}. The orbit lies between the range of validity of the the adiabatic approximation (\ref{eq:Z-AA}) and the estimate via the disk's surface density, Eq.~(\ref{eq:Z-zeta}), none of them being valid (as we can also see by the dashed and dotted curves). 
 Right: Zoom of the upper part of the orbit, showing its envelope.
  We see that Eq. (\ref{eq:Z-Sigma}) is the best approximation among the three proposed formulae for the envelopes.
 }
 \label{fig:teste2}
\end{figure*}

\begin{figure*}
\includegraphics[width=\columnwidth]{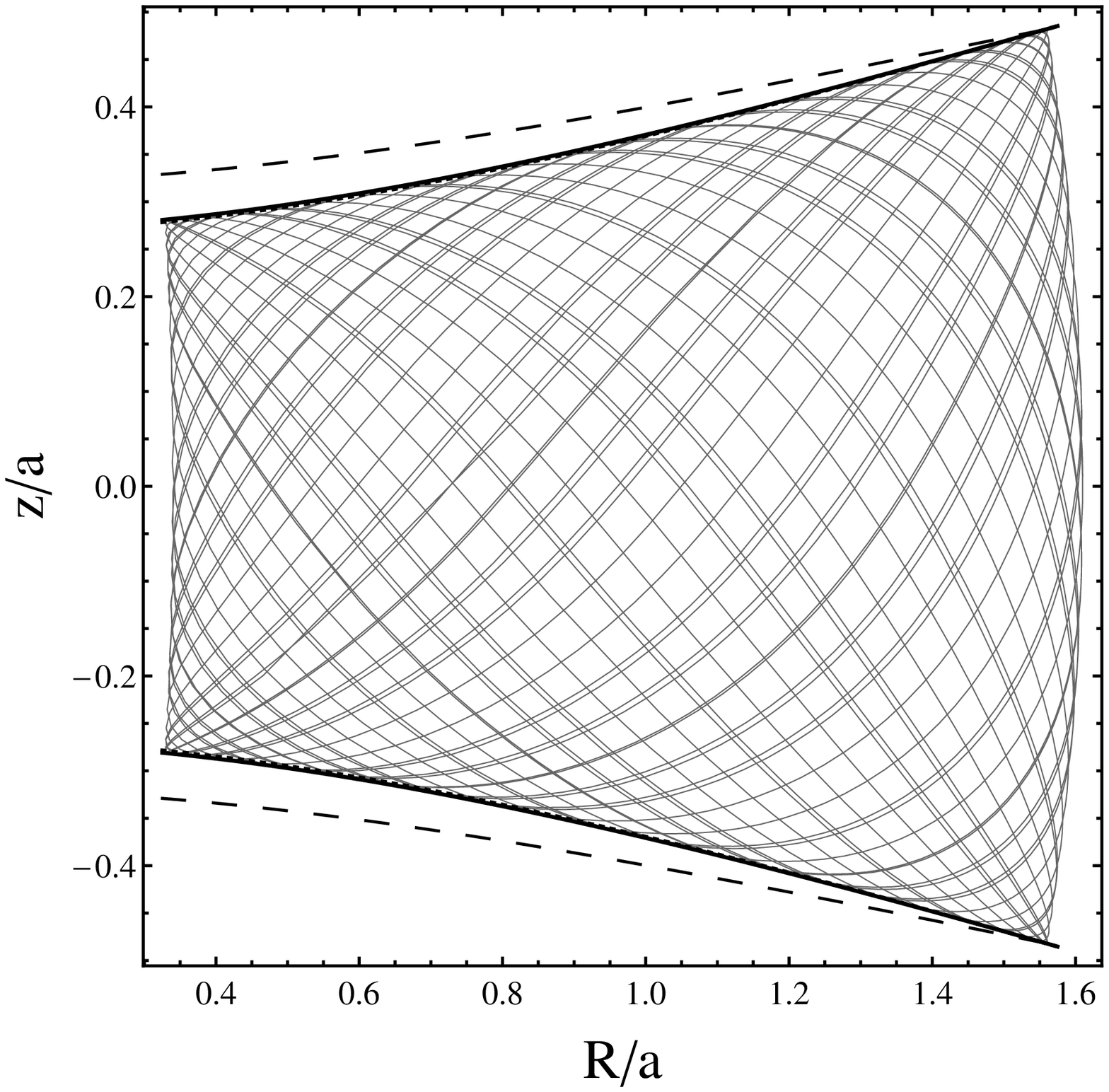}
\includegraphics[width=\columnwidth]{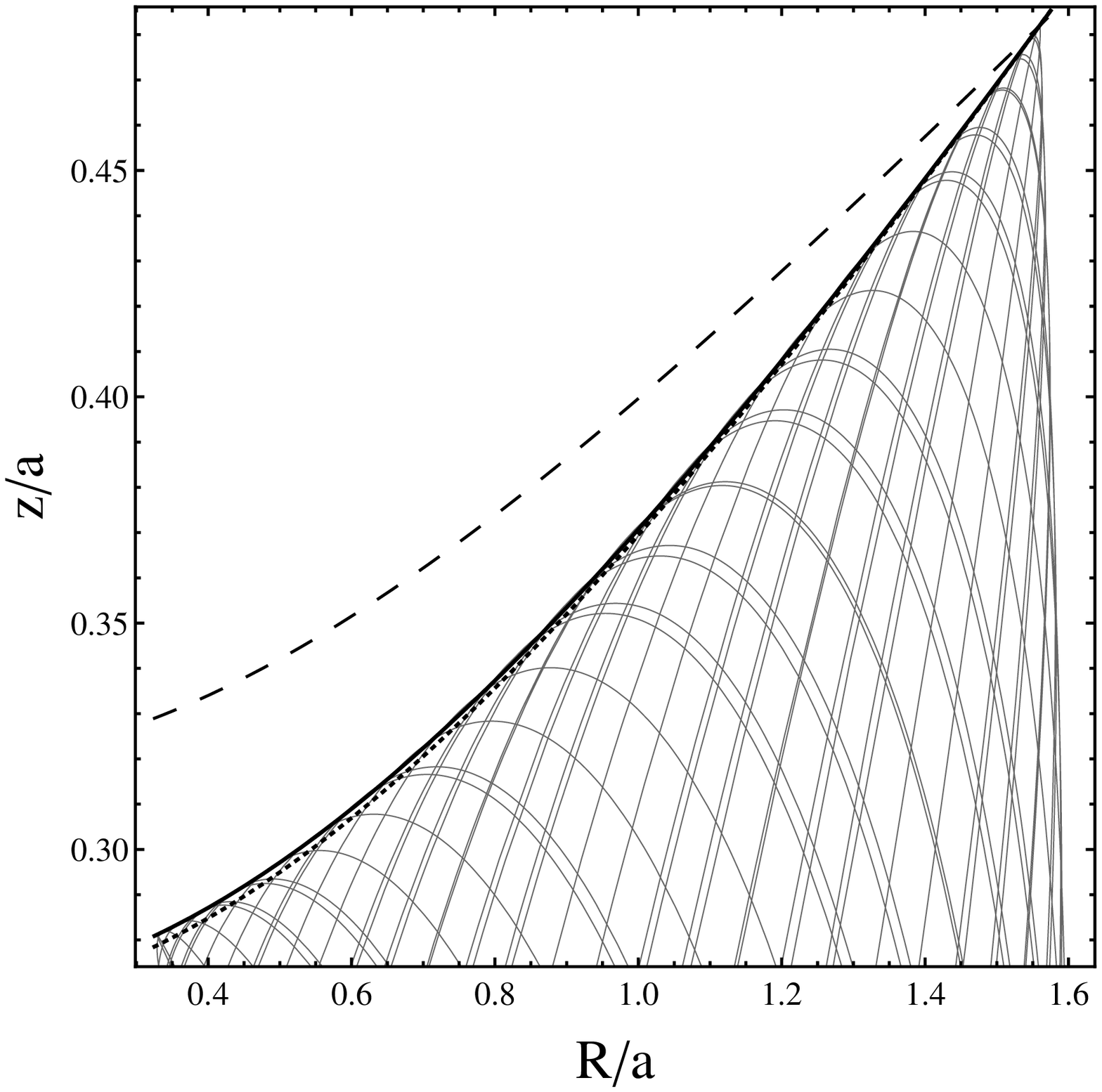}
 \caption{Left: Orbit in the Miyamoto-Nagai disk with $b/a=1/10$ (we have $\zeta/a\approx0.3$). Here, $L_z/\sqrt{GMa}=0.8$ and $Ea/GM=-0.045$. 
 Moreover, $R_0/a=1.61$, $z_0/a=1.1\times10^{-2}$, $P_{R0}=0$. Labels are as in Fig. \ref{fig:teste1}. The orbit lies in the range of validity of the envelope calculated from the disk's surface density, Eq. (\ref{eq:Z-zeta}). 
  Right: Zoom of the upper part of the orbit, showing its envelope.
  We see that Eq.~(\ref{eq:Z-Sigma}) is a slightly better approximation than the prediction from the disk's surface density (\ref{eq:Z-zeta}).
 }
 \label{fig:teste3}
\end{figure*}

We see in Figs.~(\ref{fig:teste1})--(\ref{fig:teste3}) representative examples of orbits with large radial extent and different vertical amplitudes (when compared with the disk thickness).  The orbits' envelopes are accurately described by Eqs.~(\ref{eq:Sigma-Z})--(\ref{eq:Z-Sigma}). If the vertical amplitudes are very small, being in the region 
where the usual adiabatic approximation (\ref{eq:Z-AA}) is valid (as in Fig.~\ref{fig:teste1}), both expressions 
(\ref{eq:Z-AA}) and (\ref{eq:Z-Sigma}) give good approximations for the envelopes, as expected from the above considerations. 
Moreover, as it can be seen on the right panel of
Fig.~\ref{fig:teste1}, our proposed envelope (\ref{eq:Z-Sigma}) gives an improvement to the adiabatic approximation, since it takes 
into account the whole integration interval for each galactocentric radius, and not only an equatorial-plane quantity (showing differences even when the range of integration is very small). 
In other words, the dependence of $\rho$ on $z$ in the integrand of (\ref{eq:Sigma-Z}) is important in the description of nearly equatorial orbits, since it gives better predictions than the equatorial-plane approximation (\ref{eq:Sigma-0-approx}).

This effect is also seen in orbits with intermediate amplitudes lying inside the disk ($Z(R)<\zeta$, see Fig.~\ref{fig:teste2}). The prediction from the orbit's surface density, Eqs.~(\ref{eq:Sigma-Z})--(\ref{eq:Z-Sigma}), gives results which are significantly better than  either the adiabatic approximation (\ref{eq:Z-AA}) or the disk's surface density estimate (\ref{eq:Z-zeta}), as illustrated on the right panel of Fig.~\ref{fig:teste2}. For amplitudes near the disk edge $\zeta=3b$, Eqs.~(\ref{eq:Sigma-Z})--(\ref{eq:Z-Sigma}) give predictions close to the disk's surface density estimate (\ref{eq:Z-zeta}), but with a small improvement, as exemplified in Fig.~\ref{fig:teste3}. As $Z$ approaches the disk edge, the difference between predictions (\ref{eq:Z-zeta}) and (\ref{eq:Z-Sigma}) 
becomes negligible. The prediction from (\ref{eq:Z-Sigma}) always lies between
the adiabatic approximation (\ref{eq:Z-AA}) and the disk's surface-density approximation (\ref{eq:Z-zeta}) for the Miyamoto-Nagai disk and it is, in general, 
a better description than both of these formulae.

The errors in the orbits's amplitudes, calculated from Eq.~(\ref{eq:Z-Sigma}) with the procedure described in Section \ref{sec:iterative},
are less than 1\% and are systematically smaller than the corresponding errors of formulae (\ref{eq:Z-AA}) and (\ref{eq:Z-zeta}) (see Figs.~\ref{fig:teste1}--\ref{fig:teste3}).
These errors are of the same order of magnitude as the errors appearing in the comparison of formula (\ref{eq:Z-zeta}) with 
flattened St\"ackel models \citep{vieiraRamoscaro2014ApJ}. Indeed, numerical experiments with St\"ackel potentials (the Kuzmin-Kutuzov potential of \citealp{dejongheDezeeuw1988ApJ}) show that the predictions of Eqs.~(\ref{eq:Sigma-Z})--(\ref{eq:Z-Sigma}) are also the best approximation for the orbits' envelopes.

Tests including a sparse spherical halo do not influence the predictions, as expected from the numerical experiments of \citet{vieiraRamoscaro2014ApJ} concerning near-edge amplitudes (although in this case we must consider in Eq.~(\ref{eq:Sigma-Z}) the total dynamical density of the system ``disk+halo''). The present expression is therefore an improvement for the envelopes' estimates of regular orbits inside the disk whenever the disk component is dominant, i.e. when the orbit is far from the central (spheroidal) bulge component.

\section{Conclusions and final remarks}
\label{sec:conclusions}

We presented in this paper an extension of the modeling of the envelopes of disk-crossing orbits on the system's meridional plane. It is based on the disk's surface density distribution,
which was previously presented for the cases of off-equatorial orbits around razor-thin disks \citep{vieiraRamoscaro2015MG13,vieiraRamosCaro2016CeMDA}
and of orbits in three-dimensional disks with vertical amplitudes comparable to the disk thickness \citep{vieiraRamoscaro2014ApJ}.
The generalization discussed here encompasses the two known limits
for the envelope estimates: The limit of very low vertical amplitudes 
(usual adiabatic approximation, Eq.~(\ref{eq:Z-AA}), see for example \citealp{binneytremaineGD})
and the case of vertical amplitudes near the disk thickness (the integrated surface density approach, Eqs.~(\ref{eq:Z-zeta}) and (\ref{eq:Sigma-I}), see also \citealp{vieiraRamoscaro2014ApJ}).

The proposed formula, Eq.~(\ref{eq:Z-Sigma}) with $\Sigma$ given by Eq.~(\ref{eq:Sigma-Z}),
is valid for regular tube orbits with arbitrary vertical amplitudes \mbox{$0<Z<\zeta$} (where $\zeta$ is the disk thickness), i.e. for arbitrary vertical amplitudes inside the disk.
All numerical experiments gave us relative errors smaller than 1\% for the predictions for $Z(R)$, which were systematically smaller than the predictions 
from the adiabatic approximation (\ref{eq:Z-AA}) and the predictions via the disk's surface density (\ref{eq:Z-zeta}).
In view of the present formalism and of previous results obtained for 
near--$\zeta$ vertical amplitudes \citep{vieiraRamoscaro2014ApJ}, 
this formula is expected to be valid for all regions in which the disk component is dominant.

It is worth noting that our formalism deals only with the envelope $Z(R)$ of the orbit, that is, with the ``$z$-limits'' of the orbit, since our approach is based on an extension of the vertical action variable close to the equatorial plane. The inner and outer ``$R$-limits'' of the orbit (the ``$R(z)$-envelopes'') could be obtained, in principle, by a procedure which generalizes the results obtained via the radial action variable on the equatorial plane. Such procedure is beyond the scope of this work.

A direct extension of the above formalism allows us to analyze the case of very flattened, three-dimensional axially symmetric disk systems in modified
theories of gravity which admit a Hamiltonian flow for test particles, $H=p^2/2+\Psi$, where $\Psi$ is a modified gravitational potential. This case was 
presented in \citet{vieiraRamoscaro2014ApJ} for orbits whose vertical amplitude is comparable to the disk thickness. In an analogous manner, 
we obtain for arbitrary vertical amplitudes that formula (\ref{eq:Z-Sigma}) is valid with $\Sigma$ given by
  \begin{equation}\label{eq:Sigma_Psi}
   \Sigma_\Psi(R)=f(R)\int_{-Z}^Z \rho(R,z)\, dz,
  \end{equation}
where $Z=Z(R)$ is the orbit's vertical amplitude at radius $R$ and $f(R)$ is the radial dependence of the modification to the potential, given by $\partial\Psi/\partial|z|\big|_{z=0}=f\,\Sigma$ 
when the disk component of the system is modelled as razor-thin.
There are two example cases which fit into this category of theories:
MOND \citep{bekensteinMilgrom1984ApJ} gives us $\Sigma_\Psi=\Sigma/\mu$, with $\Sigma$ given by Eq.~(\ref{eq:Sigma-Z}) and $\mu$
is MOND's interpolating function, 
and RGGR \citep{rodriguesShapiroLetelier2010JCAP} gives us $\Sigma_\Psi=[1-V_\infty^2/\Phi_N(R,0)]\,\Sigma$, where $\Sigma$ is given by Eq.~(\ref{eq:Sigma-Z}), $V_\infty$ is the asymptotic 
circular velocity of test particles on the equatorial plane, and $\Phi_N$ is the corresponding Newtonian potential. 
More details about this formalism and about the case of near-edge vertical amplitudes can be found at \citet{vieiraRamoscaro2014ApJ}.

The problem of finding the approximate third integral of motion (in terms of phase-space coordinates) which gives the envelopes (\ref{eq:Z-Sigma}) is a much more difficult task. Although the corresponding formula was recently found for orbits crossing razor-thin disks \citep{vieiraRamosCaro2016CeMDA}, this integral for three-dimensional disks would have to involve the orbit's surface density (\ref{eq:Sigma-Z}), which is a function of $R$ and $z$ (and not only on $R$, as it is in the razor-thin case). Moreover, it is not clear how the kinetic and the surface-density dependent terms would behave in the intermediate regions between small and high $z$. This question is still open and deserves a further study.






\section*{Acknowledgments}
We thank Alberto Saa for fruitful discussions. This study was financed in part by the Coordena\c{c}\~ao de Aperfei\c{c}oamento de Pessoal de N\'ivel Superior - Brasil (CAPES) - Finance Code 001.
RSSV acknowledges the support of CAPES and of S\~{a}o Paulo Research Foundation (FAPESP), grant 2010/00487-9. The authors acknowledge the support of FAPESP grant 2013/09357-9.













\bsp	
\label{lastpage}
\end{document}